%% file: draft.tex
\documentclass[aps,prl,floatfix,twocolumn,superscriptaddress,showpacs,amsmath,amssymb,showpacs,10pt]{revtex4-1}
\usepackage{graphicx}
\usepackage{epstopdf}
\usepackage{epsfig}
\usepackage{epsf}
\usepackage{url}
\usepackage[USenglish]{babel}
\usepackage{hyperref}
\def\bcen{\begin{center}}
\def\ecen{\end{center}}
\allowdisplaybreaks
\usepackage{verbatim}
\usepackage{natbib}
\usepackage{amsmath, nccmath}
\usepackage[export]{adjustbox}
\usepackage{dcolumn}
\usepackage{bm}
\usepackage{bbm}
\usepackage{lipsum}
\usepackage{color}
\usepackage{newtxtext}
\usepackage{newtxmath}
\usepackage{sidecap}

\usepackage[section]{placeins}

\begin{document}
	
	\title{Memory-Efficient Nonequilibrium Green's Function Framework Built On Quantics Tensor Trains}
	
	\author{Maksymilian \'Sroda}
	\affiliation{Department of Physics, University of Fribourg, 1700 Fribourg, Switzerland}
	\author{Ken Inayoshi}
	\affiliation{Department of Physics, Saitama University, Saitama 338-8570, Japan}
	\author{Hiroshi Shinaoka}
        \affiliation{Department of Physics, Saitama University, Saitama 338-8570, Japan}
	\author{Philipp Werner}
	\affiliation{Department of Physics, University of Fribourg, 1700 Fribourg, Switzerland}

\begin{abstract} 
One of the challenges in diagrammatic simulations of nonequilibrium phenomena in lattice models is the large memory demand for storing momentum-dependent two-time correlation functions. This problem can be overcome with the recently introduced quantics tensor train (QTT) representation of multivariable functions. Here, we demonstrate nonequilibrium Green's function simulations within the $GW$ and Migdal approximations with high momentum resolution, up to times which exceed the capabilities of standard implementations and are long enough to study, e.g., transient Floquet physics during multi-cycle electric field pulses and thermalization dynamics. The self-consistent calculation on the three-leg Kadanoff-Baym contour is fully self-contained, employing only QTT-compressed functions and input functions which are either generated directly in QTT form or obtained via quantics tensor cross interpolation.
\end{abstract}

\maketitle
	
{\it Introduction.} 
The study of nonequilibrium phenomena in correlated lattice systems is an active research field, driven by time-resolved experiments on correlated materials \cite{Giannetti2016}, the discovery of interesting nonequilibrium phases \cite{Kaneko2019,Li2020}, and novel theoretical concepts like prethermalization \cite{Berges2004,Moeckel2008} or nonthermal criticality \cite{Berges2008,Tsuji2013}. Since analytical calculations are challenging, numerical simulations play an important role in the analysis of nonequilibrium effects. But they also face severe problems, like the growth of entanglement in density-matrix renormalization group (DMRG) calculations \cite{Daley2004,White2004} 
or dynamical sign problems in Monte-Carlo-based techniques \cite{Muehlbacher2008,Werner2009}. While nonequilibrium Green's function (NEGF) methods \cite{stefanucci_van_leeuwen_2013} do not suffer from the above issues, their major bottleneck 
is the memory demand for storing momentum-dependent two-time correlation functions. Even for simple diagrammatic 
approaches, like second-order lattice perturbation theory \cite{Murray2024}, the $GW$ method \cite{Golez2016,Bittner2021}, the fluctuation-exchange approximation \cite{Stahl2021}, or the two-particle self-consistent approach \cite{Simard2022}, this has limited previous lattice simulations to short times and coarse momentum grids. 

There exist various strategies to address the memory bottleneck. 
One option is to apply approximations, e.g., the generalized Kadanoff-Baym ansatz \cite{Lipavsky1986,Hermanns2012,Kalvova2019,Tuovinen2020,Murakami2020,Schlunzen2020,Joost2020,Karlsson2021,Pavlyukh2022a,Pavlyukh2022b,Pavlyukh2022c,Tuovinen2023,Pavlyukh2024} or quantum kinetic equations \cite{Kremp1997,Stark2013,Wais2018,Queisser2019,Picano2021}, whose validity is often difficult to assess. An alternative approach is the memory-time truncation of the self-energy \cite{Schueler2018,Stahl2022}, which involves a control parameter (the cutoff time). The corresponding efficiency gains are however problem-specific, and the approach was successful mainly in the simulation of insulators. The full two-time information can be retained if one uses nonequidistant time discretizations \cite{Zwolak2008} with adaptive time stepping \cite{Meirinhos2022,Blommel2024}, but the drawback is a growing complexity of the implementation.

A recent idea that avoids the above disadvantages is to compress the nonequilibrium Green's functions \cite{Kaye2021}. A low-memory format of particular promise is the quantics-tensor-train (QTT) representation \cite{Shinaoka2023}. This approach employs a binary encoding to map the functions onto tensor trains, thus representing a grid with exponentially fine spacing, $e^{-R}$, by a linear number, $R$, of three-way tensors. Furthermore, the QTT format offers the unique possibility of compressing not only the time but also the momentum or orbital information. Previous studies showed that the bond dimension needed to accurately represent the full two-time information is moderate \cite{Shinaoka2023}, resulting in high compression ratios, and that small-scale nonequilibrium calculations are feasible \cite{Murray2024}. However, these proof-of-concept demonstrations were not only inefficient, reaching times of only a few inverse hoppings, but also not fully implemented with QTTs (relying on the matrix storage format in the input preparation), which compromised the quantics' potential for exponentially fine resolution.

Here, we develop the QTT-NEGF method into an efficient and self-contained framework and demonstrate large-scale self-consistent diagrammatic nonequilibrium simulations of the square-lattice Hubbard and Holstein models based on the widely-used $GW$ and Migdal approximations. Quenched and electric-field-driven systems are simulated up to $t_\mathrm{max}=250$ inverse hopping times on a grid of $64\times 64$ $\mathbf{k}$-points, not achievable with a conventional implementation. Furthermore, we show that in physically relevant scenarios the bond dimension only weakly depends on $t_\mathrm{max}$ and lattice size. These results will inspire the application of QTTs not only to NEGFs but to generic memory-demanding integral or integro-differential problems.

{\it Model.}
The Hamiltonian for the half-filled Hubbard model on a $N_k \times N_k$ square lattice reads
$H(t) = \sum_{\mathbf{k}\sigma} \epsilon_{\mathbf{k}-\mathbf{A} (t)} c^\dagger_{\mathbf{k}\sigma} c_{\mathbf{k}\sigma} + U(t) \sum_i (n_{i\uparrow}-\tfrac12)(n_{i\downarrow}-\tfrac12)$. 
Here, $c^\dagger_{\mathbf{k}\sigma}$ creates an electron with momentum $\mathbf{k}$ and spin $\sigma$, $\epsilon_{\mathbf{k}}=-2t_\mathrm{h}(\cos{k_x}+\cos{k_y})$ is the electron dispersion with the hopping amplitude $t_\mathrm{h}$, and $n_{i\sigma}$ is the particle-number operator for spin $\sigma$ and site $i$. The system is excited either through a time-dependent interaction $U(t)$ or an electric field $\mathbf{E}(t)$. The latter shifts the dispersion $\epsilon_{\mathbf{k}}\to\epsilon_{\mathbf{k}-\mathbf{A} (t)}$ by the vector potential $\mathbf{A}(t) = - \int_0^t \mathbf{E}(\bar{t})d\bar{t}$ \cite{Peierls1933}, corresponding to a gauge with vanishing scalar potential. We set $t_\mathrm{h}=1$, express time in units of $1/t_\mathrm{h}$, and consider paramagnetic states, suppressing the spin index hereafter.

While in the main text we focus on the Hubbard model solved within the $GW$ approximation, our methodology is applicable to a broad range of many-body systems treatable by NEGFs. Natural targets are systems with many degrees of freedom, e.g., multiband electron systems or electron-phonon problems. To demonstrate a different setup, we show results for the Holstein model in the End Matter (EM).

\begin{figure}[t]
	\centering
	\includegraphics[width=3.39in]{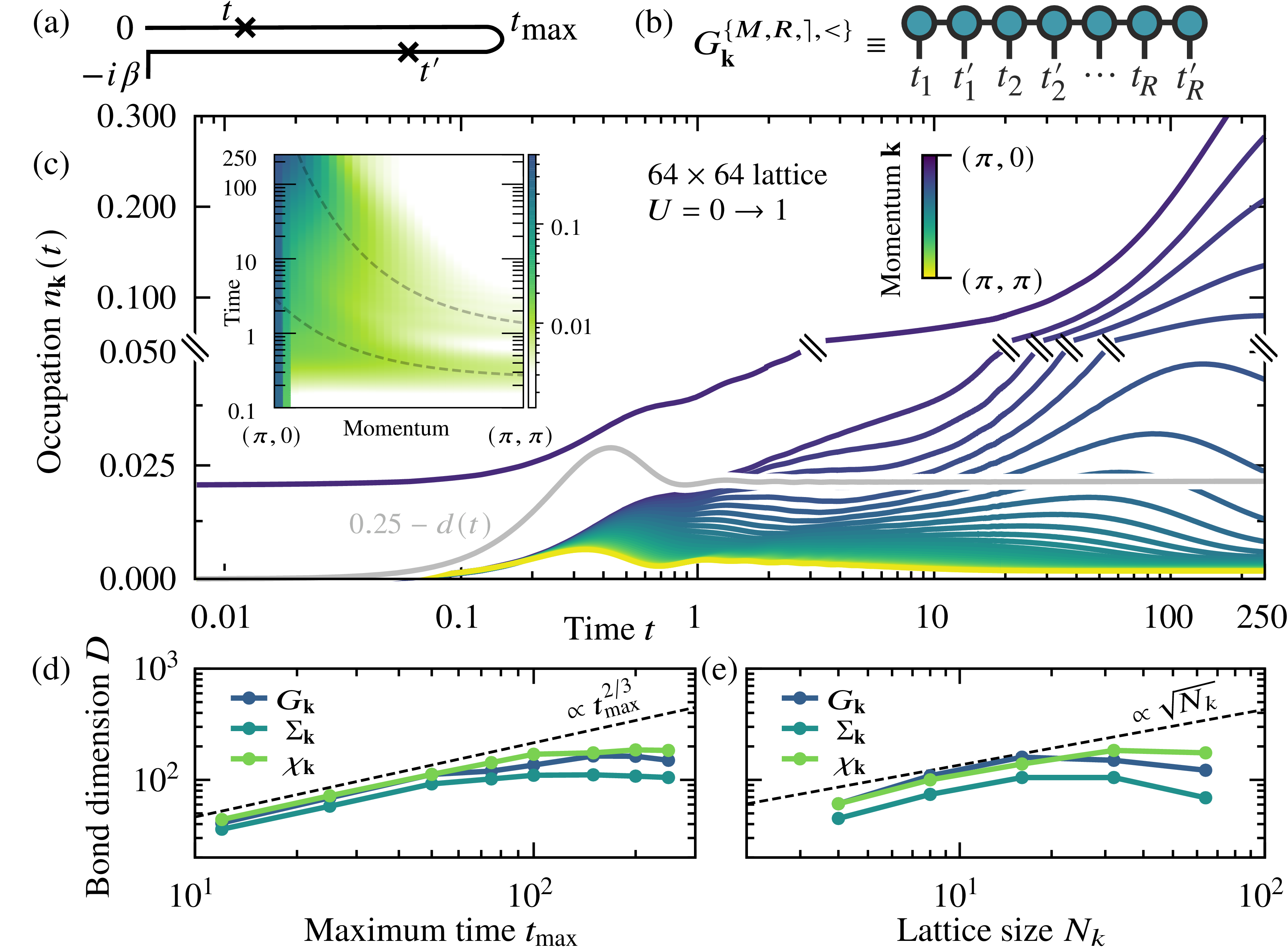}
	\caption{(a) Kadanoff-Baym contour. (b) QTT representing a Green's function component. Same length scales are ordered as neighbors \cite{Shinaoka2023}. (c) Occupations $n_\mathbf{k}(t)$ for a $U=0\to 1$ quench and $N_k=64$, $\beta=100$, where $\mathbf{k}=(\pi,\pi)\to(\pi,0)$ [with $(\pi,0)$, $(\pi,2\pi/N_k)$ not shown in the main panel]. The inset shows the same data as a color map, with the dashed lines guiding the eye. The gray line in the main panel plots the change in the double occupation $d(t)$ with respect to the initial value $0.25$. (d) Maximum bond dimension $D$ across all components and $\mathbf{k}$ points vs $t_\mathrm{max}$. Here, $N_k=32$. Usually, the lesser component contributes the largest $D$. (e) The same as in (d) but as a function of linear lattice size $N_k$ with $t_\mathrm{max}=250$. All calculations use $\epsilon_\mathrm{cutoff}=10^{-11}$ during the self-consistency loop.}
	\label{fig:quench}
\end{figure}

{\it Green's function compression.}
To study the dynamics, we calculate the interacting contour-ordered Green's function $G_\mathbf{k}(t,t')=-i\langle \mathcal{T}_\mathcal{C} c_\mathbf{k}(t)c_\mathbf{k}^\dagger(t') \rangle$, where $\mathcal{T}_\mathcal{C}$ is the time-ordering operator on the Kadanoff-Baym contour $\mathcal{C}$ [\includegraphics[height=.7em]{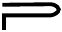}, Fig.~\ref{fig:quench}(a)], which runs along $0 \to t_\text{max} \to 0 \to -i\beta$ with the maximum time $t_\text{max}$ and inverse temperature $\beta$ \cite{Aoki2014}.  The Green's function is parametrized by 4 components: the Matsubara $G^\mathrm{M}_\mathbf{k}$, the retarded $G^\mathrm{R}_\mathbf{k}$, the left-mixing $G^\rceil_\mathbf{k}$, and the lesser $G^<_\mathbf{k}$ component. Conventional methods \cite{stefanucci_van_leeuwen_2013,Schueler2020} represent each of these components as a matrix, after discretization of the contour. Here, instead, each component is a QTT.

Conceptually, the mapping to a QTT proceeds by writing the integers enumerating the discrete time points $t$, $t'$ as binary numbers $(t_1,\ldots,t_R)_2$, $(t'_1,\ldots,t'_R)_2$.
$(0,\ldots,0)_2$ represents the first grid point, $t=0$, whereas $(1,\ldots,1)_2$ the last, $t=t_\mathrm{max}$. The matrix is thus reshaped into a 
$2R$-way tensor, which is then factorized into a tensor train---a product of $2R$ three-way tensors [Fig. \ref{fig:quench}(b)]. Extending the length $R$ of the binary string by one adds only two tensors to the train but doubles the size of the underlying matrix along both dimensions. Each tensor hence describes an exponential length scale.
The ``leg'' indices of the tensors are the binary digits of dimension 2, whereas the dimensions of the auxiliary ``bond'' indices, $D$, control both the memory demand and the faithfulness of the factorization. 
Intuitively, $D$ measures how strongly ``entangled'' two scales are \cite{Shinaoka2023,Lindsey2024}. 
Its value is set by requiring that the squared Frobenius-norm error at each
tensor of the train is less than some cutoff parameter, $|A-\tilde{A}|^2_\mathrm{F}/|A|^2_\mathrm{F} < \epsilon_\mathrm{cutoff}$ ($A$, $\tilde{A}$ mark the accurate and truncated tensors). This gives roughly a $\sqrt{\epsilon_\mathrm{cutoff}}$ precision for individual matrix elements. In practice, we never perform the above mapping, but work with compressed objects from the start.

We store $G_\mathbf{k}(t,t')$ for each $\mathbf{k}$-point as 4 QTTs, corresponding to the 4 components. We discretize the contour on an equidistant grid with spacing $dt$ on the real- and $d\tau$ on the imaginary-time axis. The grid spacing can be very fine, $dt, d\tau \lesssim 10^{-6}$, since the bond dimension of the QTT does not increase once the grid is fine enough to resolve all relevant scales. Decompressing such a $\mathbf{k}$-dependent Green's function would use $80.9$ exabytes just for the retarded and lesser components ($t_\mathrm{max}=250$, $N_k=64$ per axis \footnote{We employ symmetries to reduce the number of used $\mathbf{k}$ points, keeping ${\sim}N_k^2/8$ momenta for the quench and ${\sim}N_k^2/2$ for the electric-field calculations.}). A more fair comparison to conventional methods, assuming $dt \sim 10^{-2}$, still yields 5.6 terabytes, exceeding the memory of a typical cluster node. In contrast, the entire QTT-compressed Green's function on the very fine grid needs only $1.1$ gigabytes. 

{\it QTT-NEGF.}
The central step in a NEGF calculation is the solution of the Dyson equation $G_\mathbf{k} = G_{0\mathbf{k}} + G_{0\mathbf{k}}*\Sigma_\mathbf{k}*G_\mathbf{k}$ for a diagrammatically represented self-energy $\Sigma_\mathbf{k}$.
The computationally demanding operations are 
evaluating the Feynman diagrams for $\Sigma_\mathbf{k}$ and the convolutions ``$*$''.
Conventionally, the former involve element-wise products and the latter matrix multiplications. Here, they become tensor-network contractions \cite{Shinaoka2023,Murray2024}, performed using the fitting algorithm \cite{Verstraete2004,Stoudenmire2010}. For details of the contraction algorithms, see Ref.~\cite{Shinaoka2023}; here, we outline their application to NEGF. 
For concreteness, we show results for the standard $GW$ approximation, described in EM, but one can easily adapt the method to other approximations, e.g., the self-consistent Migdal approximation (also considered in the EM). 

A generic QTT-NEGF simulation proceeds as follows. 

(i) We prepare the noninteracting $G_{0\mathbf{k}}$ by constructing the 4 QTTs for each $\mathbf{k}$ on the whole contour. In the case of interaction quenches, $\epsilon_\mathbf{k}$ is time-independent and $G_{0\mathbf{k}}$ becomes an exponential function (with discontinuity for $G_{0\mathbf{k}}^R$ and $G_{0\mathbf{k}}^M$). Such functions can be represented as QTTs with bond dimension $D = O(1)$ for any contour 
\cite{Shinaoka2023}. In the case of electric-field simulations, we resort to the tensor cross interpolation algorithm \cite{Oseledets2011,Dolgov2020,Fernandez2022,Ritter2024,Fernandez2024}, which prepares the QTT directly by sampling a small number of suitably chosen time points; here, typically $D \sim O(100)$. 

(ii) We calculate the self-energy $\Sigma_{ij}(t,t')$ in real space in parallel and transform it back to momentum space, $\Sigma_{ij}\to \Sigma_\mathbf{k}$.

(iii) 
We solve the Dyson equation 
at each $\mathbf{k}$ in parallel. To this end, we reformulate it in terms of 4 Kadanoff-Baym equations and recast those in the form of a linear problem $AX=B$. In short-hand notation, we get $(1-G_0^M\Sigma^M)G^M=G_0^M$ for the Matsubara and $(1-G_0^R\Sigma^R)G^\alpha=B^\alpha$ for the real-time components ($\alpha=R,\rceil,<$), with $B^\alpha$ a constant term, see the Supplemental Material (SM) \cite{supp}. These linear systems are solved by a DMRG-like algorithm, with $O(D^4)$ scaling, that sequentially optimizes two tensors at a time.

Steps (ii) and (iii) are iterated until self-consistency; further details can be found in the EM.

{\it Interaction quench.}
As a first application, we consider the momentum-dependent relaxation of the Hubbard model 
after an interaction quench $U=0 \to 1$. The system is expected to show a fast relaxation to a prethermalized state, followed by slow thermalization, as found in infinite dimensions \cite{Moeckel2008,Moeckel2009,Eckstein2009,Stark2013}. We analyze this problem in two dimensions on a $64 \times 64$ lattice and up to time $t_\mathrm{max}=250$.

Figure \ref{fig:quench}(c) presents the momentum occupations $n_\mathbf{k}(t) = -i G_\mathbf{k}^<(t,t)$ along the direction $(\pi,\pi)\to(\pi,0)$. The sudden quench corresponds to a perturbation with wide frequency spectrum, and hence excites electrons up to the highest energies. This high-energy population then relaxes back towards the Fermi level in two waves: a fast one, $t \lesssim 1$, and a slow one, $t \lesssim 100$, with a plateau in between [see the inset of Fig.~\ref{fig:quench}(c)]. 
During the first wave, a part of the excess kinetic energy is exchanged into interaction energy, but in the second wave these energies are approximately separately conserved. This is shown in gray by the double occupation $d(t)$ (observables defined in SM \cite{supp}), which appears to have relaxed to its thermal value already for $t>1$. The first wave is thus associated with the prethermalization of momentum-averaged quantities, and the second with the actual thermalization. Since potential energy is essentially conserved at late times, the thermalization process converts the high kinetic energy of electrons near $\mathbf{k}=(\pi,\pi)$ into the kinetic energy of many low-energy electronic excitations \cite{Moeckel2008}. This leads to a higher-temperature distribution of the final equilibrium state. 

The possibility of resolving even longer times depends in particular on how the bond dimension $D$, which determines the computational effort, scales with the simulation time $t_\mathrm{max}$. In Fig.~\ref{fig:quench}(d), we show $D$ at the last iteration of quench calculations, where it is set 
only by the truncation cutoff $\epsilon_\mathrm{cutoff}$. $D$ initially grows as ${\sim}t_\mathrm{max}^{2/3}$ but levels off for $t_\mathrm{max} > 100$.
This growth of the bond dimension is not related to entanglement growth within the wave function, as in DMRG, but should be understood by the fact that for larger $t_\mathrm{max}$ the Green's function contains more characteristic features, similarly to how it is harder to compress a complex image rather than a plain one.
Our most expensive operations, solving the Dyson equation and computing the diagrams, scale as $O(D^4)$ in computation time and $O(D^3)$ in memory. If the $t_\mathrm{max}^{2/3}$ trend would continue, this would imply a cost $O(t_\mathrm{max}^{2.66})$ and $O(t_\mathrm{max}^{2})$. This is close to that of the conventional methods with $O(t_\mathrm{max}^3)$ and $O(t_\mathrm{max}^2)$ scaling, albeit in the QTT case likely with smaller prefactors due to the high compression ratios. However, if $D(t_\mathrm{max}\to\infty)$ saturates, as the data suggests, the computational cost for a fixed number of iterations saturates. 
In reality, while our current implementation outcompetes conventional methods, 
the number of iterations needed for convergence grows with $t_\text{max}$, and the iteration can become unstable. 
In the conclusions section, we mention how these issues could be mitigated in the future.

\begin{figure}[b]
	\centering
	\includegraphics[]{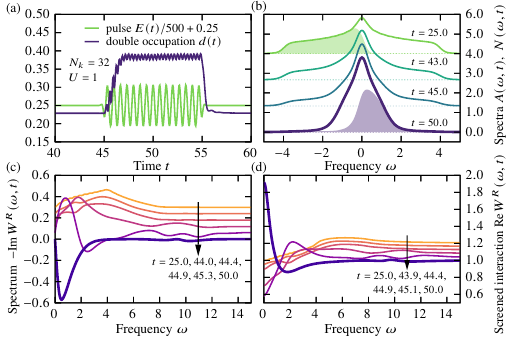}
	\caption{(a) Electric field $E(t)$ ($\Omega=3\pi$, $E_0=3\Omega$) and double occupation $d(t)$. (b) Time dependence of the fermionic spectrum $A(\omega,t)$ (lines) and the occupation $N(\omega,t)$ (shaded regions). (c),(d) Time dependence of $-\mathrm{Im}W^R(\omega,t)$ and $\mathrm{Re}W^R(\omega,t)$. The curves in panel (b) are offset by 1.33, in (c) by 0.06 and in (d) by 0.04. The parameters are: $U=1$, $N_k=32$, $\beta=10$. $G_{0\mathbf{k}}$ is prepared by tensor cross interpolation with the maximum-norm tolerance $10^{-5}$ and then truncated with $\epsilon_\mathrm{cutoff}=10^{-8}$. In the self-consistent loop, $\epsilon_\mathrm{cutoff}=10^{-11}$, $D_\mathrm{max}=140$.
	}
	\label{fig:efield}
\end{figure}

\begin{SCfigure*}[][]
	\centering
	\includegraphics[]{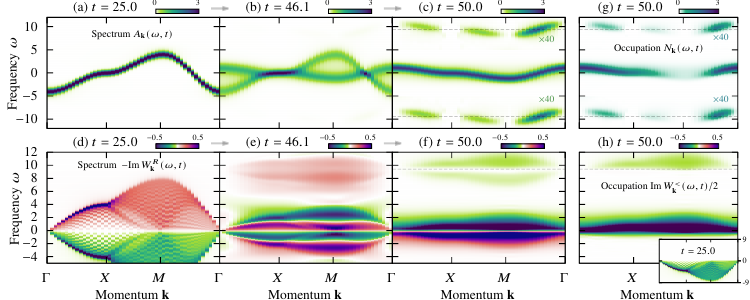}
	\caption{Evolution of (a)-(c) the fermionic spectrum $A_\mathbf{k}(\omega,t)$ and (d)-(f) the bosonic spectrum $-\mathrm{Im}W^R_\mathbf{k}(\omega,t)$ during an electric-field pulse. (g),(h) Fermionic and bosonic occupations, $N_\mathbf{k}(\omega,t)$ and $\mathrm{Im}W^<_\mathbf{k}(\omega,t)/2$. The dashed lines in panels (c),(f),(g),(h) mark $\pm \Omega=\pm 3\pi$, where the Floquet subbands appear [in (c),(g), their intensity is multiplied by $40$]. The inset in panel (h) plots $\mathrm{Im}W^<_\mathbf{k}(\omega,t)/2$ before the pulse. The parameters are the same as in Fig.~\ref{fig:efield} and the gauge with pure vector potential is used.}
	\label{fig:kresolved}
\end{SCfigure*}

Another relevant scaling for lattice calculations is the bond dimension $D$ vs the lattice size $N_k$. As shown in Fig.~\ref{fig:quench}(e), $D$ initially grows roughly as $\sqrt{N_k}$ and slows down for $N_k>32$. This suggests that future implementations of QTT-NEGF methods could benefit from compressing also the $\mathbf{k}$ dependence, potentially allowing to treat exponentially fine $\mathbf{k}$ grids. 
Furthermore, while the maximum $D$ for $G_\mathbf{k}$ in Fig.~\ref{fig:quench}(c) is 122, 
$D$ averaged over $\mathbf{k}$ is only 8 for the left-mixing, 18 for the retarded, and 50 for the lesser component. While the situation is less favorable for $\Sigma_\mathbf{k}$, $\chi_\mathbf{k}$, or in real space, this still provides a gain over conventional methods, which have the same cost for each $\mathbf{k}$. 

{\it Electric-field pulse.}
As a second application, 
we study the dynamics induced by an electric-field pulse.
The field is polarized along the diagonal direction, $\mathbf{E}(t) = E(t)(1,1)$, and we choose a box-like pulse $E(t) = E_0 \, s(t-t_\mathrm{p}) \cos\left(\Omega(t-t_\mathrm{p})\right)$ with envelope $s(t)=1/\left((1+e^{10(t-\delta/2)})(1+e^{-10(t+\delta/2)})\right)$. $E_0$ is the amplitude, $\Omega$ the frequency, $\delta$ the width, and $t_\mathrm{p}$ the pulse center. 
A continuous ac field renormalizes the hopping amplitude of a noninteracting system as $t_\mathrm{h} \to \mathcal{J}_0(E_0/\Omega)\,t_\mathrm{h}$ \cite{Dunlap1986,Holthaus1992,Tsuji2008}, with $\mathcal{J}_0$ the zeroth-order Bessel function. In a nonequilibrium situation, where the field is suddenly switched on, this can lead to an inversion of the band structure with a negative-temperature distribution, effectively switching the interaction from repulsive to attractive \cite{Tsuji2011}.
To study this behavior, we use a pulse of ${\sim}15$ cycles, 
a $32 \times 32$ lattice and maximum time $t_\mathrm{max}=100$.

Figure~\ref{fig:efield}(a) presents the evolution of the double occupation $d(t)$ during the pulse. For a repulsively interacting half-filled system $d\le 0.25$, while for an attractive interaction $d\ge 0.25$. The data thus imply that the pulse indeed inverts the sign of the interaction, which originates in the population inversion, as shown in Fig.~\ref{fig:efield}(b) by the occupation $N(\omega,t)$ \cite{supp,Freericks2009}. This panel also illustrates the band narrowing effect of the multi-cycle pulse, which here renormalizes the bands by $\mathcal{J}_0(3) \simeq -0.26$, effectively increasing the interaction strength four-fold. 

The inverted population influences the screening properties. This is shown in Fig.~\ref{fig:efield}(c), which plots the evolution of $-\text{Im}W^R(\omega,t)$ \cite{supp} during the pulse. The main peak shifts to lower frequencies, again due to the bandwidth renormalization. At $t=50$, where the inverted population develops, the bosonic spectrum changes sign, indicating an antiscreening response [see also $\text{Re}W^R(\omega,t)$ in Fig.~\ref{fig:efield}(d)], in agreement with Ref.~\cite{Golez2017}.
In contrast to the dynamical-mean-field-theory description which neglects nonlocal correlations \cite{Tsuji2011}, the $GW$ description of the population-inverted driven state captures two effects enhancing the correlation strength: the bandwidth narrowing and the antiscreening.

The high momentum and frequency resolution of the QTT method enables a more detailed study of the above physics. Figure~\ref{fig:kresolved} shows the evolution of the momentum-resolved fermionic and bosonic spectra, $A_\mathbf{k}(\omega,t)$ and $-\mathrm{Im}W^R_\mathbf{k}(\omega,t)$ \cite{supp}. Before the pulse, the spectra embody the well-known equilibrium dispersions of weakly interacting fermions on a square lattice [Fig.~\ref{fig:kresolved}(a),(d)]. At the pulse edge, Fig.~\ref{fig:kresolved}(b), a flipped fermionic band starts forming, leading to a complicated transient state with two dispersions seemingly present (the probe pulse has a width $\delta=2$, see SM \cite{supp}). This is also reflected in the bosonic spectrum [Fig.~\ref{fig:kresolved}(e)]. Later, during the pulse, the fermionic band completely flips [Fig.~\ref{fig:kresolved}(c)], and its population is inverted [Fig.~\ref{fig:kresolved}(g)]. The bosonic subsystem also evolves into an inverted state, indicated by the sign change of the low-energy contribution to $-\mathrm{Im}W^R_\mathbf{k}(\omega,t)$ [Fig.~\ref{fig:kresolved}(f)]. This inversion is directly visible in the bosonic occupation [Fig.~\ref{fig:kresolved}(h)]. Here, the low-energy contribution shifts from negative (inset) to positive frequencies (main panel). Furthermore, in both the fermionic and bosonic spectra, one can recognize Floquet subbands \cite{Tsuji2008} at integer multiples of the pulse frequency, $n\Omega=n3\pi$, whose occupation is also inverted [Fig.~\ref{fig:kresolved}(g),(h)]. After the pulse, these effects disappear and the system thermalizes at an increased temperature.

Electric-field calculations are more demanding than interaction quenches. Here, the entire calculation uses a fixed $D_\mathrm{max}$, slowly increased during convergence up to $D_\mathrm{max}=140$. Unlike in the quench setup, almost all $\mathbf{k}$ points have $D$ saturated to this value. 
We noticed, however, that $D_\mathrm{max}\sim 140$ is sufficient to reach the same convergence precision at both $t_\mathrm{max}=50,100$. This suggests that in the present study, where the system relaxes quickly after the pulse, it is the pulse itself that controls the maximum $D$.
Partitioning the contour into several QTTs is thus likely to improve the efficiency, as only the subdomain containing the pulse would have a substantial $D$.

{\it Conclusions.}
We presented a NEGF simulation framework based on QTTs and demonstrated its potential with Hubbard and Holstein model simulations on the square lattice. The largest lattices and times we reached were $64 \times 64$ and $t_\mathrm{max}=250$, which would be prohibitively expensive for conventional methods. We also showed evidence that the bond dimension $D$, which controls the memory demand, saturates with growing $t_\mathrm{max}$ and lattice size. It thus appears feasible to simulate even larger systems and longer times with improved implementations, and to essentially overcome the 
memory bottleneck in lattice simulations.

The main obstacle in the current implementation is the large number of iterations needed for convergence with globally updated Kadanoff-Baym equations and their instability.
A promising solution are patching schemes \cite{Shinaoka2023}, wherein the NEGFs are partitioned over several small tensor trains defined on subdomains of the contour. Not only does this allow to grow the contour as the calculation proceeds, utilizing causality, but it should further improve the $D(t_\mathrm{max})$ scaling. Such a partitioning is a generalization of the hierarchical method of Ref.~\cite{Kaye2021} and at least the memory scaling $O(t_\mathrm{max}\log t_\mathrm{max})$ should be achievable. Judging by the saturation observed here, it could be even better. We will explore causality-based solutions of the KBE using QTT patching schemes in upcoming papers. Since the quantics representation utilizes exponentially fine grids and in principle allows to compress also the $\mathbf{k}$-dependence, it is a variant of the hierarchical scheme that promises highly economical and efficient long-time simulations. We expect that applications with new physical insights will follow in future works making use of our methodology. 

\begin{acknowledgments}
{\it Acknowledgements.}
We thank A.\ Kauch, M.\ Ritter, and X.\ Waintal for helpful discussions. M.\'S.\ and P.W.\ acknowledge support from SNSF Grant No.\ 200021-196966. K.I.\ was supported by JSPS KAKENHI Grant No.~23KJ0883, Japan.
H.S.\ was supported by JSPS KAKENHI Grants No.~22KK0226, and No.~23H03817 as well as JST FOREST Grant No.~JPMJFR2232 and JST PRESTO Grant No.~JPMJPR2012, Japan.
The calculations were performed on the beo05 cluster at the University of Fribourg. The QTT-NEGF implementation is written in Julia \cite{bezanson2017julia} and is based on the ITensor library \cite{itensor} and libraries developed by the tensor4all group \cite{Fernandez2024,tensor4all}.

{\it Data availability.} The data that support the findings of this article are openly available \cite{data}.
\end{acknowledgments}

\bibliography{ref}

\appendix*
\onecolumngrid
\section{End Matter}
\twocolumngrid

{\it Details on the $GW$ implementation.}
The self-consistent $GW$ approximation is defined via the real-space self-energy 
$$\Sigma_{ij}(t,t') = i G_{ij}(t,t')\delta W_{ij}(t,t'),$$ 
where $G_{ij}(t,t')=(1/N_k^2) \sum_\mathbf{k} e^{-i(\mathbf{r}_i-\mathbf{r}_j)\cdot\mathbf{k}} G_\mathbf{k}(t,t')$ and $\delta W_{ij}(t,t')$ is the dynamical part of the screened interaction. The full interaction is given by $W_{ij}(t,t')=U_{ij}(t,t')+\delta W_{ij}(t,t')$, where $U_{ij}(t,t')=U(t)\delta_{ij}\delta_\mathcal{C}(t,t')$ is the bare interaction with $\delta_\mathcal{C}(t,t')$ the contour delta function. 
We calculate $\delta W_{ij}(t,t')$ from the charge susceptibility $\chi_{ij}(t,t')$ as $\delta W_{ij}(t,t')=U(t)\chi_{ij}(t,t')U(t')$. 
The susceptibility is in turn obtained from the bosonic Dyson equation (in integral form)
$$\chi_\mathbf{k}(t,t') = P_\mathbf{k}(t,t') + [P_\mathbf{k}*U_\mathbf{k}*\chi_\mathbf{k}](t,t'),$$ 
where $*$ denotes the contour convolution \cite{Aoki2014} and 
$$P_{ij}(t,t') = -iG_{ij}(t,t')G_{ji}(t',t)$$ 
is the polarization bubble. 

The interacting Green's function is computed from the fermionic Dyson equation
$$G_\mathbf{k}(t,t') = G_{0\mathbf{k}}(t,t') + [G_{0\mathbf{k}}*\Sigma_\mathbf{k}*G_\mathbf{k}](t,t').$$ 
Here, 
$$G_{0\mathbf{k}}(t,t') = -i[\theta_\mathcal{C}(t,t')-f_\beta(\epsilon_\mathbf{k}(0))]e^{-i\int_{t'}^t d\bar{t}\epsilon_\mathbf{k}(\bar{t})}$$ 
is the noninteracting Green's function \cite{Freericks2005}, with $f_\beta(\omega)$ the Fermi function and $\theta_\mathcal{C}(t,t')$ the contour step function \cite{Aoki2014}.

In the main text, we discussed a generic QTT-NEGF self-consistency loop. More specifically, our $GW$ implementation proceeds as follows. (i) First, we prepare the noninteracting $G_{0\mathbf{k}}$. Then, we use it to calculate the polarization $P_{0\mathbf{k}}$ [see step (iv)], which constitutes the ansatz for the charge susceptibility, $\chi_{0\mathbf{k}} = P_{0\mathbf{k}}$. (ii) We calculate $\Sigma_\mathbf{k}$ by Fourier transforming $\chi_{\mathbf{k}}\to \chi_{ij}$ to obtain the real-space $\delta W_{ij}$. With this, we compute $\Sigma_{ij}$ and transform the result back to momentum space, $\Sigma_{ij}\to \Sigma_\mathbf{k}$. (iii) We solve the fermionic Dyson equation at each $\mathbf{k}$ to update $G_\mathbf{k}$. (iv) We calculate the polarization $P_\mathbf{k}$ by Fourier transforming $G_\mathbf{k}\to G_{ij}$, evaluating the bubble, and transforming the result back, $P_{ij}\to P_\mathbf{k}$. (v) Finally, we solve the bosonic Dyson equation updating $\chi_\mathbf{k}$.

Steps (ii)-(v) are iterated until self-consistency, i.e., until each (normalized) Green's function component differs between subsequent iterations by ${\lesssim} 10^{-4}$-$10^{-3}$ in terms of the Frobenius norm (see SM \cite{supp} for a convergence analysis and a benchmark against reference data). 
We first converge the equations on the Matsubara, and then on the real-time axis. To reduce the computational effort, we restrict the maximum bond dimension $D_\mathrm{max}$ in the first iterations and increase it slowly during convergence. 
In the quench calculations, 
the accuracy is controlled in late iterations via a fixed $\epsilon_\mathrm{cutoff}$. 
To stabilize the equations, we also slowly ramp up the interaction $U$ within preprocessing iterations.
Although these iterations are cheap, to avoid a divergence or stagnation of the global update of the Kadanoff-Baym equations, the ramp needs to be very slow for larger $U > 2$, necessitating an excessive number of extra iterations.
This problem will be solved in a future study by a causality-based updating scheme.

Note that there are 4 Fourier transformations per iteration. The rationale behind this is to take advantage of the diagonality of the Dyson equation in $\mathbf{k}$-space and the locality of the diagrams in real-space by performing the appropriate tensor-network contractions in parallel. To speed up the Fourier transforms, we 
implemented a simple realization of the Cooley-Tukey fast Fourier transform \cite{Cooley1965}, which works on an $N_k \times N_k$ matrix of QTTs, and evaluates QTT sums in the butterfly. 

{\it Results for an electron-phonon problem.}
In the main text we focused on the Hubbard model. However, the proposed QTT method is not restricted to this setup, but should be applicable to generic many-body systems treatable by the nonequilibrium Green's function formalism. To demonstrate this, in this section, we present additional results for an electron-phonon problem.

\begin{figure}[t]
	\centering
	\includegraphics[]{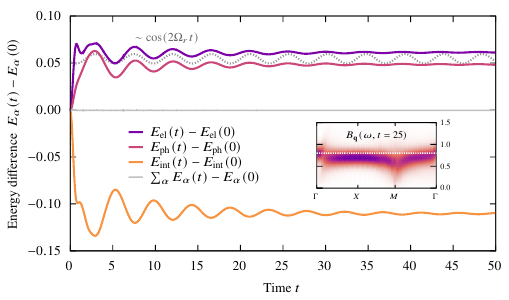}
	\caption{Electron-phonon coupling $g$ quench in the electron-phonon model \eqref{eq:holstein} with $N_k=32$ and $\beta=20$. The main panel shows the change in the electron, phonon, and interaction energies after a sudden $g=0 \to 0.5$ quench. The sum of the energy differences is shown in gray. The dotted dark gray line shows $a\cos(2\Omega_r t)+b$ with $a$, $b$ constants and $\Omega_r\approx 0.69$ the renormalized phonon frequency estimated from the position of the maximum of the local phonon spectral function $B(\omega,t)$. The inset shows the momentum-resolved phonon spectral function $B_\mathbf{q}(\omega,t=25)$ with the dotted white line marking the bare phonon frequency $\Omega=0.8$.}
	\label{fig:holstein}
\end{figure}

We consider the half-filled Holstein Hamiltonian 
\begin{equation}
    \begin{aligned}
        H(t) =&~H_\mathrm{el} + H_\mathrm{ph} + H_\mathrm{int}(t)\\[.5em]
        =& \sum_{\mathbf{k}\sigma} \epsilon_{\mathbf{k}} c^\dagger_{\mathbf{k}\sigma} c_{\mathbf{k}\sigma} + \Omega\!\sum_{\mathbf{q}} b^\dagger_{\mathbf{q}} b_{\mathbf{q}} + g(t)\!\sum_i (b_i^\dagger + b_i)(n_i-1).
    \end{aligned}\label{eq:holstein}
\end{equation}
Here, $c^\dagger_{\mathbf{k}\sigma}$ creates an electron with momentum $\mathbf{k}$ and spin $\sigma$, $\epsilon_{\mathbf{k}}=-2t_\mathrm{h}(\cos{k_x}+\cos{k_y})$ is the electron dispersion, $b^\dagger_{\mathbf{q}}$ creates an optical phonon with momentum $\mathbf{q}$ and frequency $\Omega$ ($b^\dagger_i$ is the corresponding creation operator in real space), whereas $n_i=\sum_\sigma c^\dagger_{i\sigma}c_{i\sigma}$ is the electron number operator on site $i$ of a two-dimensional square lattice. The system is excited through a time-dependent modulation of the electron-phonon coupling $g(t)$. As in the main text, the energy unit is the hopping amplitude $t_\mathrm{h}=1$, the time unit is $1/t_\mathrm{h}$, and we consider paramagnetic states. The model is solved within the renormalized Migdal approximation \cite{Migdal1958,Murakami2015,Abdurazakov2018}, which includes the feedback of the electrons on the phonon subsystem and is conserving.

Figure~\ref{fig:holstein} presents the energy dynamics following a sudden coupling quench $g(t=0)=0\to g(t>0)=0.5$. As discussed in previous studies using a $\mathbf{k}$-independent self-energy approximation \cite{Murakami2015,Abdurazakov2018}, after a quench, such a momentum-averaged quantity is expected to oscillate with twice the renormalized phonon frequency $2\Omega_r$. This is indeed observed in Fig.~\ref{fig:holstein}, where $\Omega_r$ is taken as the position of the peak in the local phonon spectral function $B(\omega,t)$. Note also that the total energy is conserved after the quench (solid gray line), which confirms the convergence of the calculation.

In the inset of Fig.~\ref{fig:holstein}, we furthermore show the momentum-resolved phonon spectral function $B_\mathbf{q}(\omega,t=25)$. The latter clearly displays a phonon renormalization and a frequency softening at $\mathbf{q}=M=(\pi,\pi)$. This phonon softening is known from equilibrium studies \cite{Dee2019,Nosarzewski2021} and is related to a charge-density-wave instability in the lattice.

The above calculations reproduce the expected physics resulting from the electron-phonon feedback in a nonequilibrium setting. Note that treating this feedback is very memory-demanding, because of the need for storing the two-time phonon correlation function for each momentum $\mathbf{q}$.

\include{supp}

\end{document}

%% file: supp.tex
%================================================================================
% Supplementary information
%================================================================================
\makeatletter
    \close@column@grid
    \clearpage
    \twocolumngrid
\makeatother
\appendix
%\starttocentries
\setcounter{page}{1}
\setcounter{figure}{0}
\setcounter{equation}{0}
\setcounter{table}{0}
\newcommand{\rom}[1]{\uppercase\expandafter{\romannumeral #1\relax}}
\renewcommand{\refname}{Supplemental References}
\renewcommand{\figurename}{Supplemental Figure}
\renewcommand{\thefigure}{S\arabic{figure}}
\renewcommand{\theHfigure}{S\arabic{figure}} % otherwise hyperrefs point to main text
\renewcommand{\citenumfont}[1]{#1}
\renewcommand{\bibnumfmt}[1]{[#1]}
\renewcommand{\thepage}{S\arabic{page}}
\renewcommand{\theequation}{S\arabic{equation}}
\renewcommand{\theHequation}{S\arabic{equation}} % otherwise hyperrefs point to main text
\renewcommand{\thetable}{S\Roman{table}}
\renewcommand{\theHtable}{S\arabic{table}} % otherwise hyperrefs point to main text
%================================================================================
\onecolumngrid

\begin{center}
{\bf \uppercase{Supplemental Material}}\\
\vspace{3pt}
{\bf \large \makeatletter\@title\makeatother}\\
\vspace{10pt}
by M. {\'S}roda, K. Inayoshi, H. Shinaoka, and P. Werner
\end{center}
\vspace{10pt}

\twocolumngrid
%--------------------------------------------------------------------------------
\section{Kadanoff-Baym equations}

The Dyson equation reads
\begin{equation}
	G(t,t') = G_{0}(t,t') + [G_{0}*\Sigma*G](t,t'),
\end{equation}
where $[a*b](t,t') = \int_\mathcal{C} d\bar{t} a(t,\bar{t}) b(\bar{t},t')$ is a convolution on the contour $\mathcal{C}$ and the spin and momentum indices are suppressed to simplify the notation. This equation can be viewed as a linear problem by rearranging it as 
\begin{equation}
	(1-G_0*\Sigma*{}\!)G = G_0.
\end{equation}
By means of the Langreth rules \cite{Langreth1976,stefanucci_van_leeuwen_2013}, one can also express it in terms of 4 Kadanoff-Baym equations for the Matsubara, retarded, left-mixing, and lesser components. These new equations involve only convolutions on the real- or imaginary-time axes, $[a\cdot b](t,t') = \int_0^{t_\mathrm{max}} d\bar{t} a(t,\bar{t}) b(\bar{t},t')$ and $[a\star b](t,t') = -i\int_0^{\beta} d\bar{t} a(t,\bar{t}) b(\bar{t},t')$, respectively. In the integral form and rearranged to reveal the linear-problem structure, they read
\begin{align}
    &(1-G_0^M\star\Sigma^M\star{})G^M = G_0^M,\label{eq:mat}\\
    &(1-G_0^R\cdot\Sigma^R\cdot{})G^R = G_0^R,\label{eq:ret}\\
    &\begin{aligned}
        (1-G_0^R\cdot\Sigma^R\cdot{})G^\rceil = G_0^\rceil &+ G_0^R \cdot \Sigma^\rceil \star G^M \\{}&+ G_0^\rceil \star \Sigma^M \star G^M,
    \end{aligned}\label{eq:tv}\\
    &\begin{aligned}
        (1-G_0^R\cdot\Sigma^R\cdot{})G^< = G_0^< &+ G_0^R \cdot \Sigma^< \cdot G^A\\
        {}&+ G_0^\rceil \star \Sigma^\lceil \cdot G^A\\
        {}&+ G_0^< \cdot \Sigma^A \cdot G^A\\
        {}&+ G_0^R \cdot \Sigma^\rceil \star G^\lceil\\
        {}&+ G_0^\rceil \star \Sigma^M \star G^\lceil,
    \end{aligned}\label{eq:les}
\end{align}
where also the usual advanced ($A$) and right-mixing ($\lceil$) components appear. These components can be reconstructed from the other 4 components by symmetry \cite{Aoki2014}. Note that the Matsubara component is defined here with an additional $i$ factor, as compared to the usual convention. That is, $G^M(t,t') = -i\langle \mathcal{T}_M c(t)c^\dagger(t')\rangle$, where both arguments lie on the imaginary-time axis and $\mathcal{T}_M$ is a corresponding time-ordering operator.
Since the Matsubara equation \eqref{eq:mat} defines the equilibrium initial state at $t<0$, it is decoupled from the rest and can be solved to self-consistency separately as a first step. The remaining equations are solved one by one in the order \eqref{eq:ret}-\eqref{eq:les} at each iteration of the self-consistent loop.

In the actual numerical implementation, the equations \eqref{eq:mat}-\eqref{eq:les} are discretized. The convolutions thus become matrix multiplications, $[a\cdot b]_{t,t'} = \sum_{\bar{t}} a_{t,\bar{t}} w_{\bar{t}} b_{\bar{t},t'}$, $[a \star b]_{t,t'} = \sum_{\bar{t}} a_{t,\bar{t}} v_{\bar{t}} b_{\bar{t},t'}$, where $w$, $v$ are diagonal matrices containing the integration weights for the real- and imaginary-time axes, respectively. We apply the trapezoidal rule $w=dt\,\mathrm{diag}(\frac{1}{2},1,\ldots,1,\frac{1}{2})$, $v=-id\tau\,\mathrm{diag}(\frac{1}{2},1,\ldots,1,\frac{1}{2})$, which should lead to negligible discretization errors for the very fine grids ($dt, d\tau \lesssim 10^{-6}$) that we use.

\section{Observables}
The double occupation is obtained from the contour convolution $\Sigma_\mathbf{k} * G_\mathbf{k}$ as
\begin{equation}
    d(t) = \frac{-i}{N_k^2} \sum_\mathbf{k} \frac{[\Sigma_\mathbf{k} * G_\mathbf{k}]^<(t,t)}{U(t)} + 0.25.
\end{equation}

The fermionic spectra $A(\omega,t)$, $N(\omega,t)$ and $A_\mathbf{k}(\omega,t)$, $N_\mathbf{k}(\omega,t)$ are obtained by the following formula for the photoemission intensity \cite{Freericks2009}
\begin{align}
    \mathcal{I}(\omega,t) ={}& \mathrm{Im}\int_0^{t_\mathrm{max}} dt_1 \int_0^{t_\mathrm{max}} dt_2\, S(t_1-t) S(t_2-t) e^{i\omega(t_1-t_2)}\nonumber
    \\ &\times \mathcal{G}(t_1, t_2). \label{eq:pes}
\end{align}
Here, $\mathcal{I}(\omega,t)$ represents the (possibly momentum-resolved) spectrum obtained from a two-time function $\mathcal{G}(t_1, t_2)$, while $S(t)=\exp{\left(-t^2/(2\delta^2)\right)}$ is the envelope of the probe pulse with duration $\delta$. We use $\mathcal{G}=G^</(2\pi),\,-G^R/\pi$ for the occupied $N(\omega,t)$ and full $A(\omega,t)$ spectra, respectively. We take $\delta=2$ for the momentum-resolved spectra and $\delta=4$ for the local spectra. The intergral transform \eqref{eq:pes} is prepared as a single tensor-train operator by viewing it as two quantum Fourier transforms, which are readily expressible as quantics tensor trains \cite{Shinaoka2023}. Note that applying the above formula results in nonnegative fermionic spectra, as physically expected.

The frequency-dependent screened interaction $W^{R,<}(\omega,t)$ [or $W^{R,<}_\mathbf{k}(\omega,t)$] is defined via the relative-time Fourier transform,
\begin{align}
    W^{R,<}(\omega, t) = \int dt_\mathrm{rel} e^{i \omega t_\mathrm{rel}} W^{R,<}(t + t_\mathrm{rel}, t - t_\mathrm{rel}),\label{eq:trelft}
\end{align}
where $t=(t_1+t_2)/2$, $t_\mathrm{rel}=(t_1-t_2)/2$ are the average and relative times, respectively. The above formula is implemented by first transforming the quantics tensor train into the new variables and then applying a quantum Fourier transform as a tensor-train operator \cite{Shinaoka2023}. Additionally, a Gaussian window with standard deviation $\sigma=10$ is used to reduce ringing artifacts.

\section{Convergence analysis}

The convergence precision of our nonequilibrium Green's function calculations is measured in terms of the Frobenius-norm difference between subsequent iterations
\begin{equation}
    \epsilon_\mathrm{conv} = \mathrm{max}_\mathbf{k} \sum_{\alpha=R,\rceil,<}\frac{|X^\alpha_{\mathbf{k},\mathrm{new}}-X_{\mathbf{k},\mathrm{old}}^{\alpha}|_\mathrm{F}}{|X_{\mathbf{k},\mathrm{old}}^\alpha|_\mathrm{F}}.\label{eq:convprec}
\end{equation}
Here, $X=G,\chi$ and it is assumed that $|X_{\mathbf{k},\mathrm{new}}^\alpha|_\mathrm{F} \approx |X_{\mathbf{k},\mathrm{old}}^\alpha|_\mathrm{F}$, which should be the case close to convergence. We take the maximum across all $\mathbf{k}$ points, so that we can monitor the precision of the most slowly converging $\mathbf{k}$ point. The Frobenius norm is used, as it is the cheapest and most straightforward norm to evaluate with a tensor train. It is also the norm used to measure the faithfulness of the compressed tensor-train representation (see the main text). We are typically converging to $\epsilon_\mathrm{conv} \sim 10^{-4}$-$10^{-3}$.

\begin{figure}[t]
	\centering
	\includegraphics[]{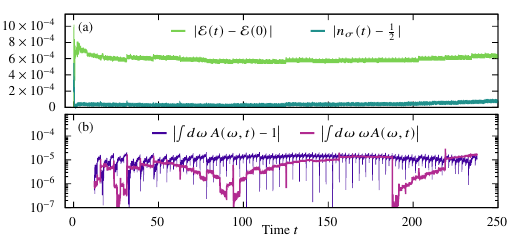}
	\caption{Convergence analysis for the quench calculations. (a) Conservation of energy $\mathcal{E}(t)$ and particle density per spin $n_\sigma(t)$. (b) Fulfillment of the fermionic sum rules for the zeroth and first moments. For the analysis of sum rules, we evaluate the fermionic spectral function $A(\omega,t)$ with Eq.~\eqref{eq:trelft} instead of Eq.~\eqref{eq:pes}. The first moment is given by $\int\!d\omega\,\omega A(\omega,t)=U(\sum_\sigma n_\sigma -1)/2$ \cite{White1991}, which evaluates to 0 at half-filling.}
	\label{fig:conv-quench}
\end{figure}

To check whether our convergence criterion is sufficiently strict, we analyze the fermionic sum rules and particle-number and energy conservation at each time instant $t$. Figure~\ref{fig:conv-quench} presents such an analysis for the quench results of Fig.~\ref{fig:quench}(c) of the main text. 
The precision $\epsilon_\mathrm{conv}$ reached in this calculation is ${\sim} 4.2 \times 10^{-4}$ for the bosonic Dyson equation and ${\sim} 2.0 \times 10^{-3}$ for the fermionic one. The conservation laws are obeyed up to a similar tolerance. The total energy is conserved up to ${\sim} 6 \times 10^{-4}$, while the particle number is conserved up to ${\sim} 1 \times 10^{-4}$ (apart from the initial spikes around $t=0$).  Furthermore, the fermionic sum rules, which are often taken as an adequate convergence check \cite{Freericks2006}, are obeyed up to a higher accuracy, ${\sim} 1 \times 10^{-5}$.
These are satisfactory results which agree with the expectation of about $3$ or $4$ significant digits of precision, sufficient for the data reported in the main text.

\begin{figure}[t]
	\centering
	\includegraphics[]{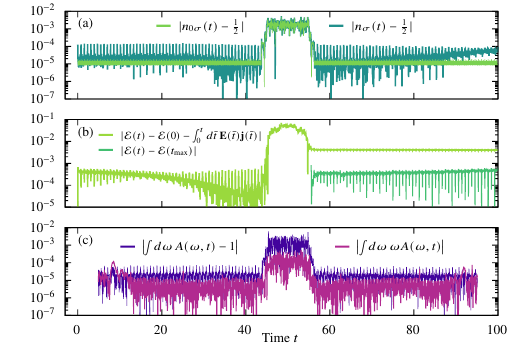}
	\caption{Convergence analysis for the electric-field calculations. (a) Conservation of the particle density per spin: $n_\sigma(t)$ is the density for an interacting system, whereas $n_{0\sigma}(t)$ is for the noninteracting one. (b) Conservation of the energy $\mathcal{E}(t)$. The electric field $\mathbf{E}(t)$ pumps energy into the system, hence the energy conservation is expressed as $\dot{\mathcal{E}}(t) = \mathbf{E}(t)\mathbf{j}(t)$, where $\mathbf{j}(t)=j(t)(1,1)$ is the current. After integration, one gets $\mathcal{E}(t)-\mathcal{E}(0) = \int_0^t d\bar{t} \,\mathbf{E}(\bar{t})\mathbf{j}(\bar{t})$, which is more convenient to check numerically. We perform the integration on the very fine grid within the quantics-tensor-train format by an appropriate tensor-network contraction. After the pulse, we also show how well the energy is conserved with respect to the final value $\mathcal{E}(t_\mathrm{max})$. (c) Fulfillment of the fermionic sum rules for the zeroth and first moments. For the analysis of sum rules, we evaluate the fermionic spectral function $A(\omega,t)$ with Eq.~\eqref{eq:trelft} instead of Eq.~\eqref{eq:pes}. The first moment is given by $\int\!d\omega\,\omega A(\omega,t)=U(\sum_\sigma n_\sigma -1)/2$ \cite{White1991}, which evaluates to 0 at half-filling.}
	\label{fig:conv-efield}
\end{figure}

Figure~\ref{fig:conv-efield} presents an analogous analysis for the electric-field results of Figs.~\ref{fig:efield} and \ref{fig:kresolved} of the main text. The calculation reached a precision ${\sim} 5.2 \times 10^{-4}$ for the bosonic Dyson equation and ${\sim} 2.9 \times 10^{-4}$ for the fermionic one. Here, the conservation laws are well obeyed before and after the pulse, with the particle density $n_\sigma(t)$ conserved up to ${\sim}10^{-4}$, the energy $\mathcal{E}(t)$ up to ${\sim}10^{-3}$, and the sum rules up to ${\sim}10^{-5}$. Although during the pulse the deviations are larger, they are still acceptable. These deviations most badly affect the $\int_0^t d\bar{t} \,\mathbf{E}(\bar{t})\mathbf{j}(\bar{t})$ integral needed to evaluate the energy conservation in Fig.~\ref{fig:conv-efield}(b). We show, however, that although the value of the energy after the pulse is somewhat inaccurate, the energy is still well conserved, and its difference with respect to the final value $\mathcal{E}(t_\mathrm{max})$ is small, $<10^{-3}$. The additional errors within the pulse duration can be traced back to an inaccuracy present already in the noninteracting $G_{0\mathbf{k}}$ (also the particle density $n_{0\sigma}(t)$ is not very well conserved during the pulse).  To fix these issues, one would likely need to impose a smaller tolerance in the tensor cross interpolation algorithm that prepares $G_{0\mathbf{k}}$, and also decrease the singular-value-decomposition cutoff used to truncate $G_{0\mathbf{k}}$ after preparation. This, however, would increase the bond dimension, which is already quite large ($D=96$) and would make the calculations more demanding. The current accuracy is in any case sufficient for the type of data presented in the main text.

The convergence of our solutions on the imaginary axis is also measured with Eq.~\eqref{eq:convprec}, but for the Matsubara component $X^M$. The Matsubara equation usually converges to high precision in very few iterations, $< 10$, and we hence do not present a detailed convergence analysis.

\section{Comparison to reference data}

In this section, we compare the results from the QTT method to reference data from the open-source package NESSi \cite{Schueler2020}, which implements a conventional matrix-based integrator for the Kadanoff-Baym equations. 
For this test, we simulate the Hubbard model, as in the main text, but in one dimension. The electron dispersion is thus given by $\epsilon_{k}=-2t_\mathrm{h}\cos{k}$, and we use $N_k=64$ momentum points. The system is perturbed by a single-cycle electric-field pulse $E(t) =  E_0\, e^{-(t-t_\mathrm{p})^2/\delta^2} \!\sin(\Omega (t-t_\mathrm{p}))$ with $E_0=5$, $\Omega=4$, $t_\mathrm{p}=2\pi/\Omega$, and $\delta=t_\mathrm{p}/\sqrt{4.6}$.

\begin{figure}[b]
	\centering
	\includegraphics[]{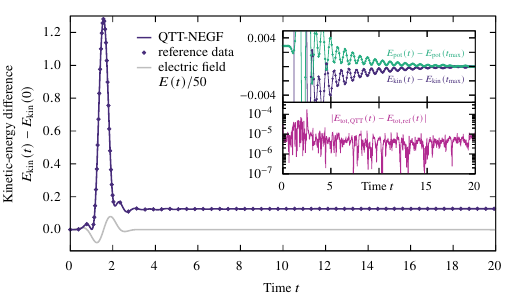}
	\caption{Comparison of a QTT-NEGF simulation result to reference NESSi data for the energy dynamics in a one-dimensional Hubbard model excited by an electric field. Main panel: time evolution of the change in the kinetic energy $E_\mathrm{kin}(t)-E_\mathrm{kin}(0)$. The gray line shows the electric-field perturbation, which is rescaled by $1/50$ to fit into the range of the plot. Top inset: Zoom on the relaxation dynamics by subtracting the long-time limit, $E_\alpha(t) - E_\alpha(t_\mathrm{max})$, where $\alpha=\mathrm{kin},\mathrm{pot}$ corresponds to the kinetic and potential energies, respectively. The lines show QTT-NEGF data, while the dots are the reference data. Bottom inset: Absolute error of the total energy in the QTT-NEGF solution with respect to the NESSi reference. Note that the total energy has a nontrivial dynamics during the pulse, since the field pumps energy into the system. The system parameters are $N_k=64$, $U=1$, $\beta=20$.}
	\label{fig:nessi}
\end{figure}

Figure~\ref{fig:nessi} compares the dynamics of the kinetic and potential energy, calculated with both implementations. The NESSi data were converged up to precision $<10^{-8}$ at each time step ($dt \simeq 0.02$), so that they can serve as a reference. The QTT-NEGF simulation used a maximum-norm tolerance of $10^{-8}$ to prepare $G_{0\mathbf{k}}$ and the subsequent tensor-network algorithms used $\epsilon_\mathrm{cutoff}=10^{-12}$, after the usual slow bond dimension ramp (see the main text). The self-consistency loop reached a convergence precision of $\epsilon_\mathrm{conv} \simeq 10^{-7}$, measured with Eq.~\eqref{eq:convprec}. The results agree perfectly between the two approaches, confirming the validity and accuracy of our QTT-based method. In particular, the absolute error in the total energy is $\lesssim 10^{-5}$ on average, consistent with the rough estimate $\sqrt{\epsilon_\mathrm{cutoff}} = 10^{-6}$ for the precision of the QTT compression.

